\documentclass[twocolumn,tighten]{aastex63}
\usepackage{amsmath}
\usepackage{threeparttable}
\usepackage{float}
\usepackage{graphicx}
\usepackage{CJK}
\usepackage{xspace}

\newcommand{\package}[1]{\tt{#1}}

\newcommand{\code}[1]{\texttt{#1}\xspace}
\newcommand{\unit}[1]{\ensuremath{\mathrm{\,#1}}\xspace}

\newcommand{\Teff}{\ensuremath{T_\mathrm{eff}}\xspace}
\newcommand{\logg}{\ensuremath{\log\,g}\xspace}

\newcommand{\kms}{\unit{km\,s^{-1}}}

\shorttitle{$r$-process in Kraken vs. GSE}
\shortauthors{Naidu et al.}

\begin{document}
\begin{CJK*}{UTF8}{gbsn}

\title{Evidence from Disrupted Halo Dwarfs that $r$-process Enrichment via Neutron Star Mergers is Delayed by $\gtrsim500$ Myrs}

\correspondingauthor{Rohan P. Naidu}
\email{rohan.naidu@cfa.harvard.edu}
\author[0000-0003-3997-5705]{Rohan P. Naidu}
\affiliation{Center for Astrophysics $|$ Harvard \& Smithsonian, 60 Garden Street, Cambridge, MA 02138, USA}
\author[0000-0002-4863-8842]{Alexander~P.~Ji}
\affiliation{Department of Astronomy \& Astrophysics, University of Chicago, 5640 S Ellis Avenue, Chicago, IL 60637, USA}
\affiliation{Kavli Institute for Cosmological Physics, University of Chicago, Chicago, IL 60637, USA}
\author[0000-0002-1590-8551]{Charlie Conroy}
\affiliation{Center for Astrophysics $|$ Harvard \& Smithsonian, 60 Garden Street, Cambridge, MA 02138, USA}
\author[0000-0002-7846-9787]{Ana Bonaca}
\affiliation{Observatories of the Carnegie Institution for Science, 813 Santa Barbara St., Pasadena, CA 91101, USA}
\author[0000-0001-5082-9536]{Yuan-Sen Ting (丁源森)}
\affiliation{Research School of Astronomy \& Astrophysics, Mount Stromlo Observatory, Cotter Road, Weston Creek, ACT 2611, Canberra, Australia}
\affiliation{Research School of Computer Science, Australian National University, Acton ACT 2601, Australia}
\author[0000-0002-5177-727X]{Dennis Zaritsky}
\affiliation{Steward Observatory, University of Arizona, 933 North Cherry Avenue, Tucson, AZ 85721-0065, USA}
\author[0000-0001-5484-4987]{Lieke ~A.~C.~van~Son}
\affiliation{Center for Astrophysics $|$ Harvard \& Smithsonian, 60 Garden Street, Cambridge, MA 02138, USA}
\author[0000-0002-4421-4962]{Floor S. Broekgaarden}
\affiliation{Center for Astrophysics $|$ Harvard \& Smithsonian, 60 Garden Street, Cambridge, MA 02138, USA}
\author[0000-0002-8224-4505]{Sandro Tacchella}
\affiliation{Department of Physics, Ulsan National Institute of Science and Technology (UNIST), Ulsan 44919, Republic of Korea}
\author[0000-0002-0572-8012]{Vedant Chandra}
\affiliation{Center for Astrophysics $|$ Harvard \& Smithsonian, 60 Garden Street, Cambridge, MA 02138, USA}
\author[0000-0003-2352-3202]{Nelson Caldwell}
\affiliation{Center for Astrophysics $|$ Harvard \& Smithsonian, 60 Garden Street, Cambridge, MA 02138, USA}
\author[0000-0002-1617-8917]{Phillip Cargile}
\affiliation{Center for Astrophysics $|$ Harvard \& Smithsonian, 60 Garden Street, Cambridge, MA 02138, USA}
\author[0000-0003-2573-9832]{Joshua S. Speagle (\begin{CJK*}{UTF8}{gbsn}沈佳士\ignorespacesafterend\end{CJK*})}
\altaffiliation{Banting \& Dunlap Fellow}
\affiliation{David A. Dunlap Department of Astronomy \& Astrophysics, University of Toronto, 50 St. George Street, Toronto ON M5S 3H4, Canada}
\affiliation{Dunlap Institute for Astronomy and Astrophysics, University of Toronto, 50 St George Street, Toronto, ON M5S 3H4, Canada}
\affiliation{Department of Statistical Sciences, University of Toronto, 100 St George St, Toronto, ON M5S 3G3, Canada}

\begin{abstract}
The astrophysical origins of $r$-process elements remain elusive. Neutron star mergers (NSMs) and special classes of core-collapse supernovae (rCCSNe) are leading candidates. Due to these channels' distinct characteristic timescales (rCCSNe: prompt, NSMs: delayed), measuring $r$-process enrichment in galaxies of similar mass, but differing star-formation durations might prove informative. Two recently discovered disrupted dwarfs in the Milky Way's stellar halo, Kraken and \textit{Gaia}-Sausage Enceladus (GSE), afford precisely this opportunity: both have $M_{\star}\approx10^{8}M_{\rm{\odot}}$, but differing star-formation durations of ${\approx}2$ Gyrs and ${\approx}3.6$ Gyrs. Here we present $R\approx50,000$ Magellan/MIKE spectroscopy for 31 stars from these systems, detecting the $r$-process element Eu in all stars. Stars from both systems have similar [Mg/H]$\approx-1$, but Kraken has a median [Eu/Mg]$\approx-0.1$ while GSE has an elevated [Eu/Mg]$\approx0.2$. With simple models we argue NSM enrichment must be delayed by $500-1000$ Myrs to produce this difference. rCCSNe must also contribute, especially at early epochs, otherwise stars formed during the delay period would be Eu-free. In this picture, rCCSNe account for $\approx50\%$ of the Eu in Kraken, $\approx25\%$ in GSE, and $\approx15\%$ in dwarfs with extended star-formation durations like Sagittarius. The inferred delay time for NSM enrichment is $10-100\times$ longer than merger delay times from stellar population synthesis -- this is not necessarily surprising because the enrichment delay includes time taken for NSM ejecta to be incorporated into subsequent generations of stars. For example, this may be due to natal kicks that result in $r$-enriched material deposited far from star-forming gas, which then takes $\approx10^{8}-10^{9}$ years to cool in these galaxies. 
\end{abstract}

\section{Introduction}
\label{sec:introduction}

Approximately half the elements in the modern periodic table originate in the rapid neutron capture process ($r$-process, see \citealt[][]{Cowan21} for a recent review). Despite this outsized importance, the astrophysical birth-sites of the $r$-process remain elusive. Neutron star mergers (NSMs) and special classes of core-collapse supernovae (rCCSNe\footnote{We use ``rCCSNe" to denote the special core-collapse supernovae that produce $r$-process elements (e.g., hypernovae, magnetorotational supernovae, magnetars), and ``CCSNe" to reference the entire population of core-collapse supernovae.}) are leading candidates. The one NSM witnessed via electromagnetic radiation (GW170817) has shown signatures of $r$-process production \citep[e.g.,][]{Kasen17,Drout17,Kasliwal17}, but NSMs alone might be unable to explain features of $r$-process chemistry observed in the Milky Way (MW) and its dwarf galaxies \citep[e.g.,][]{Cote19,Haynes19,Reichert20,Skuladottir20,Tsujimoto21}. On the other hand, rCCSNe such as magnetorotational hypernovae have been theorized as feasible channels, but empirical evidence remains tentative \citep[e.g.,][]{Ting12,Nishimura17,Halevi18,Siegel19,Kobayashi20,Yong21}.

Here we seek new constraints on the $r$-process from the stellar halo. The key development in the last few years, enabled by the \textit{Gaia} mission, has been the identification of the distinct dwarf galaxies whose debris constitutes the halo \citep[e.g.,][]{Belokurov18,Helmi18,Myeong19,Koppelman19,Yuan20,Kruijssen20,Naidu20, Horta21}. Like some of the MW's surviving dwarfs, some disrupted dwarfs are also chemical fossils -- they formed all their stars in the first few Gyrs of the Universe before they were tidally disrupted by the Galaxy. The most ancient surviving dwarfs have been studied for decades to isolate the imprints of the $r$-process since they are enriched only by a few generations of star-formation \citep[e.g.,][]{Ji16,Duggan2018,Skuladottir19}. Now we turn to the ancient disrupted dwarfs, whose unique characteristics are particularly suited to unraveling the $r$-process.

Of particular importance, debris from multiple $M_{\rm{\star}}\gtrsim10^{8} M_{\rm{\odot}}$ systems has been identified within a few kpc from the Sun (e.g., from the \textit{Gaia}-Sausage Enceladus galaxy accreted at redshift $z\approx2$, \citealt[][]{Helmi18, Belokurov20,Naidu21}). In this mass regime, stochastic effects due to the expected rarity of production sites (e.g., roughly one NSM expected for $\approx10^{5} M_{\rm{\odot}}$, e.g., \citealt[][]{Ji16}) and undersampling of the initial mass function (e.g., \citealt[][]{Koch08}) do not complicate the interpretation of $r$-process chemistry unlike in low mass dwarf galaxies. Accounting for shallow potential wells that are unable to hold on to enriched gas is also no longer an issue (e.g., GSE had a total mass $\approx2\times10^{11}M_{\rm{\odot}}$, \citealt[][]{Naidu21}).

A practical advantage is that some disrupted dwarfs are the ``nearest" dwarf galaxies, since their debris circulates through the Solar neighborhood and is easily within the grasp of high-resolution spectroscopy. Low and high-resolution spectra for tens of thousands of GSE stars have already been acquired, making it the most spectroscopically studied dwarf galaxy of all time \citep[e.g.,][]{RAVE2, Bird19,Conroy19b,Mackereth20,Buder21} though only a small fraction of these spectra are currently able to support $r$-process studies due to the signal-to-noise, resolution, and wavelength coverage required \citep[e.g.,][]{Aguado20,Matsuno21}.

Another useful feature of the disrupted dwarfs is that many independent chemo-dynamical methods may be deployed to determine their star-formation (SF) duration, which is crucial in disentangling ``delayed" sources of enrichment like NSMs from ``prompt" sources like rCCSNe. By SF duration we mean the period over which a galaxy has assembled the bulk of its stellar mass. SF durations are being inferred from diverse methods such as: age distributions from spectra of hundreds of \textit{individual} main-sequence turnoff stars \citep[e.g.,][]{Bonaca20}, color magnitude diagram fitting \citep[e.g.,][]{Gallart19}, tailored merger simulations \citep[e.g.,][]{Naidu21}, orbital signatures \citep[e.g.,][]{Pfeffer20}, and ages/metallicities of accompanying GC systems \citep[e.g.,][]{Kruijssen19}.

In this paper we constrain the channels of $r$-process enrichment by contrasting GSE and Kraken, two of the most massive dwarfs to have merged with the Milky Way. These galaxies have similar stellar and halo masses, but different star-formation durations, thereby affording a controlled experiment to disentangle the $r$-process contributions of rCCSNe and NSMs. We select GSE and Kraken samples from APOGEE DR16 (\S\ref{sec:sampleselection}), describe our Magellan/MIKE followup (\S\ref{sec:mike}), discuss the mass and star-formation durations of GSE and Kraken (\S\ref{sec:massandsfh}), present the MIKE abundances (\S\ref{sec:results}), interpret them with simple chemical evolution models (\S\ref{sec:models}), and close by discussing the implications of our results (\S\ref{sec:discussion}). We use $r_{\rm{gal}}$ to denote the total Galactocentric distance, $E_{\rm{tot}}$ for the total orbital energy, and $e$ for the eccentricity of orbits. We adopt a \citet[][]{Planck18} cosmology to convert redshifts into lookback times.

\begin{figure*}
\centering
\includegraphics[width=\linewidth]{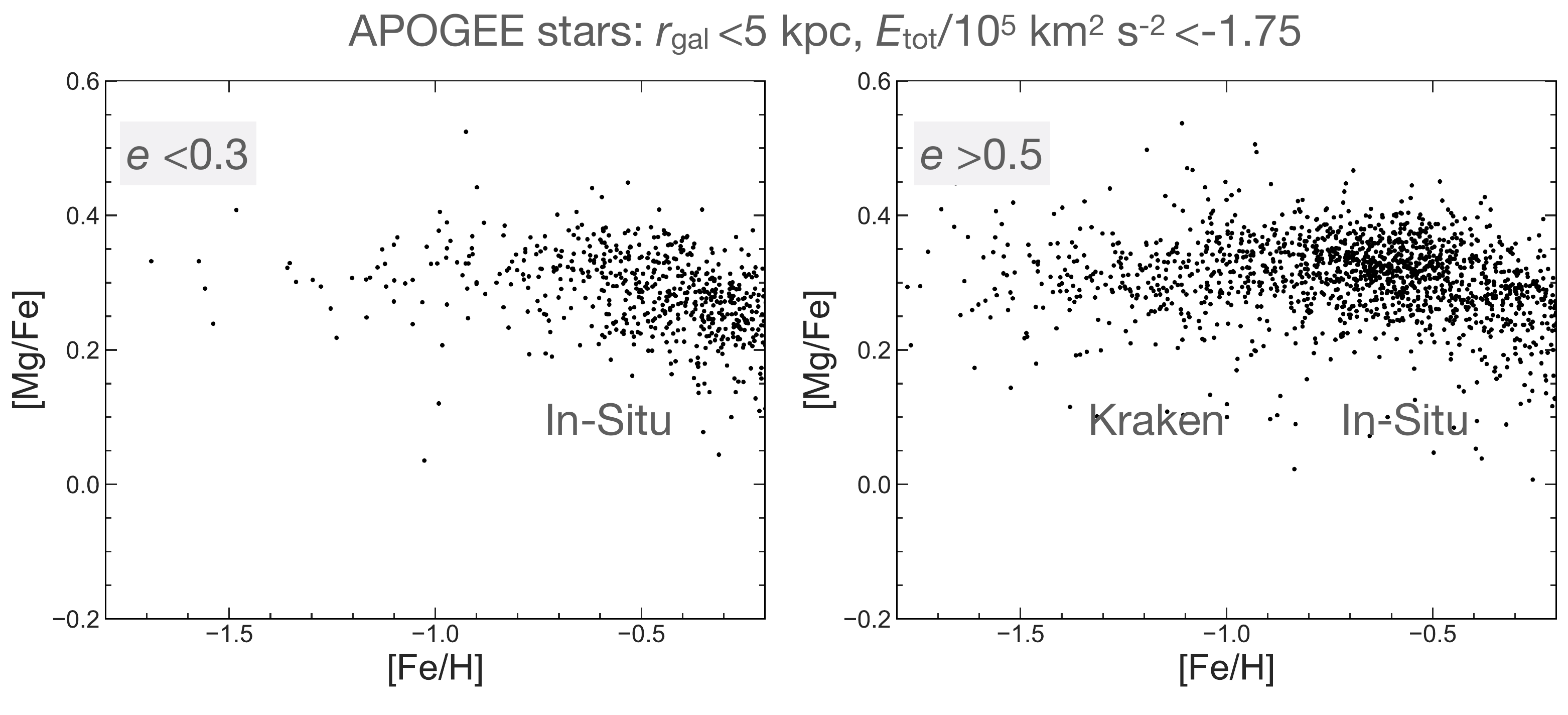}
\caption{Revealing Kraken in the APOGEE dataset. [Mg/Fe] vs. [Fe/H] for APOGEE stars in the inner 5 kpc of the Galaxy selected to have low energy (energy cut from Fig. \ref{fig:krakensel}). The eccentric stars (right panel) show a metal-poor, Mg-enhanced population that is absent at lower eccentricities (left panel). This preferential concentration at $e>0.5$ tracks the Kraken GCs and is expected from dynamical friction considerations for a high mass-ratio merger \citep[e.g.,][]{Naidu21}.}
\label{fig:ecc}
\end{figure*}

\begin{figure*}
\centering
\includegraphics[width=\linewidth]{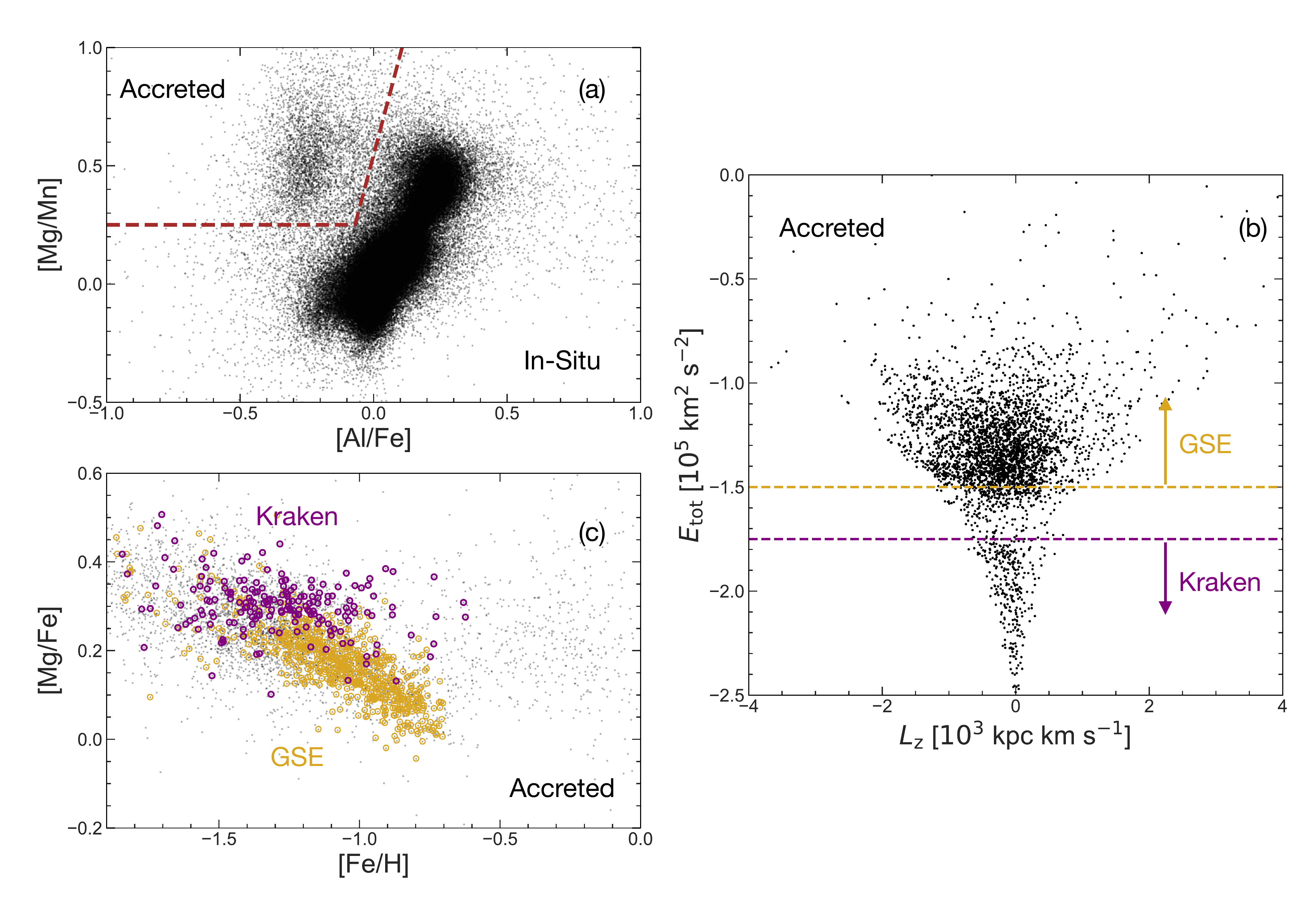}
\caption{Selecting Kraken and GSE samples. \textbf{Panel (a):} All APOGEE stars that satisfy our quality cuts are shown in the [Mg/Mn] vs. [Al/Fe] chemical plane that has been used to separate accreted stars from in-situ stars \citep[e.g.,][]{Das20}. We use the brown dashed line to demarcate the two visibly distinct populations. \textbf{Panel (b):} Total orbital energy vs. the $z$-component of angular momentum for accreted stars identified via Panel (a). There is a clear gap around $E_{\rm{tot}}\approx-1.6\times 10^{5}$ km$^{2}$ s$^{-2}$ between a high energy population (GSE) and a low energy population (Kraken). \textbf{Panel (c):} The final selected Kraken (purple) and GSE (golden) samples (see Eqns. \ref{eqn:kraken}, \ref{eqn:gse}) show tight, coherent [Mg/Fe] vs. [Fe/H] sequences characteristic of distinct dwarfs. The Kraken sequence is $\alpha$-enhanced with no visible ``knee", which is expected for a galaxy disrupted earlier, and forming stars more efficiently than GSE.}
\label{fig:krakensel}
\end{figure*}

\section{Sample Selection}
\label{sec:sampleselection}

The samples studied here were selected from APOGEE DR16 \citep[][]{Jonsson20} cross-matched with \textit{Gaia} EDR3 \citep[][]{GaiaEDR3}. We adopt data quality cuts for APOGEE following \citet{Horta21} and \textit{Gaia} EDR3 following \citet{GaiaEDR3}. Globular cluster (GC) member stars are excluded based on radial velocity and proper motion cuts -- $3\sigma$ around the mean cluster radial velocity and proper motions for stars lying within $3\times$ the tidal radius of the cluster on-sky. GC parameters are sourced from \citet{Baumgardt18,Baumgardt19,Vasiliev21}. Distances to stars beyond $\approx3$ kpc from the Sun, especially towards the Galactic center, are uncertain based on \textit{Gaia} parallaxes alone so we rely on data-driven distances fit to these stars from \citet{Leung19} and an updated version of \citet{Hogg19,Eilers19} based on DR16 and \textit{Gaia} EDR3 (A.C. Eilers, private comm.), discarding stars for which these authors' distances disagree by $>1\sigma$. With full 6D phase-space coordinates from these datasets, we are able to compute dynamical quantities of interest (angular momenta, energies, eccentricities) following \citet{Naidu20}. We use these quantities along with APOGEE abundances to design our Kraken and GSE selections.

\subsection{Kraken}

From the clustering of a dozen GCs in the age-metallicity plane as well as various dynamical planes, it has been inferred that the remains of a massive ($M_{\rm{\star}}\gtrsim10^{8}M_{\rm{\odot}}$) dwarf galaxy (``Kraken") lies buried in the inner few kpc of the MW \citep[e.g.,][]{Kruijssen19,Massari19,Forbes20, Pfeffer21}. The sheer number of GCs, and the fact that they are confined to the inner regions of the Galaxy (apocenters $\lesssim5$ kpc) has been interpreted as evidence for an early accretion event ($z\gtrsim2$) that was likely the MW's highest mass-ratio merger \citep{Kruijssen20}. 

Almost all GCs associated with Kraken in the age-metallicty plane \citep[e.g.,][]{Kruijssen19} lie at $r_{\rm{gal}}\lesssim5$ kpc and are eccentric ($\langle e \rangle \approx 0.75$), as expected for a massive merger wherein the infalling galaxy is rapidly radialized due to efficient dynamical friction \citep[e.g.,][]{Amorisco17, Koppelman20b,Naidu21,Vasiliev21DF}. Supporting the existence of Kraken, even low-energy field stars at $<5$ kpc show a metal-poor sequence that preferentially occurs on radial orbits ($e>0.5$, Figure \ref{fig:ecc}). Importantly, this population does not define a continuous distribution in eccentricity, and preferentially occurs at $e>0.5$. This distribution mirrors the Kraken GCs. Further, it is evidence that we are not observing an in-situ population like the splashed disk that has a continuous eccentricity distribution \citep[e.g.,][]{Bonaca20, Belokurov20}.

Empirically, the [Mg/Mn] vs. [Al/Fe] plane has been shown to separate accreted stars from in-situ MW stars \citep{Hawkins15,Das20,Horta21}. In Figure \ref{fig:krakensel} we select accreted stars based on this plane and find these stars separate into two clear subsets -- one at high energy, consistent with GSE \citep[e.g.,][]{Naidu20} and another at low energy, exactly where the Kraken GCs are found to cluster \citep[e.g.,][]{Massari19}. Further, the low-energy accreted stars define a coherent sequence in the [Fe/H] vs. [Mg/Fe] plane characteristic of a single dwarf galaxy that is distinct from GSE. The elevated [Mg/Fe]$\approx0.3$ of this sequence is consistent with a galaxy disrupted before Type Ia supernovae began dominating chemical evolution.

In summary, the chemodynamical signatures of the Kraken dwarf galaxy inferred from GCs are closely mirrored by a population of stars in the inner Galaxy that we select as follows:
\begin{equation}
\begin{aligned}
\label{eqn:kraken}
(r_{\rm{gal}}/[\rm{kpc}]<5) \land (\textit{e}>0.5)\\
\land\ \rm{[Mg/Mn]}>0.25\\
\land\ \rm{[Mg/Mn]}-4.25\rm{[Al/Fe]}>0.55\\
\land\ E_{\rm{tot}}/[10^{5}\ \rm{km}^{2}\ \rm{s}^{-2}]<-1.75.
\end{aligned}
\end{equation}

We end this subsection by noting that Kraken has been identified under other names in the literature -- the host of the ``low-energy" GCs \citep[][]{Massari19}, ``Koala" \citep[][]{Forbes20}, and ``Inner Galaxy Structure"/``Heracles" \citep[][]{Horta21}. Given the close correspondence in the properties of our selected stars and the \citet[][]{Kruijssen20} Kraken GCs, we are confident our criteria select the stellar debris of the Kraken dwarf galaxy.

\subsection{Gaia-Sausage Enceladus (GSE)}

To select a pure GSE sample we focus on eccentric, accreted stars as identified in the [Mg/Mn] vs. [Al/Fe] plane at energies and distances larger than Kraken motivated by Figure \ref{fig:krakensel}:

\begin{equation}
\label{eqn:gse}
\begin{aligned}
(r_{\rm{gal}}/[\rm{kpc}]>5) \land\ (\textit{e}>0.7)\\
\land\ \rm{[Mg/Mn]}>0.25\\
\land\ \rm{[Mg/Mn]}-4.25\rm{[Al/Fe]}>0.55\\
\land\ E_{\rm{tot}}/[10^{5}\ \rm{km}^{2}\ \rm{s}^{-2}]>-1.50.
\end{aligned}
\end{equation}

These criteria are very similar in spirit to typical literature selections of GSE (see \citealt[][]{Buder21} for a compilation). The median abundances ([Fe/H] and [Mg/Fe]) are in excellent agreement with \citet{Naidu20}, who did not make cuts in the [Mg/Mn] vs. [Al/Fe] plane. Further note that the \citet{Naidu20} sample measures metallicities down to [Fe/H]$\approx-3.0$, and so the agreement in medians is an important check that the metallicity distribution function (MDF) sampled by APOGEE is not meaningfully distorted by the lack of coverage at [Fe/H]$\lesssim-2.0$.

\section{Observations and Abundance Analysis}
\label{sec:mike}
We observed 20 Kraken stars and 11 GSE stars with Magellan/MIKE \citep{Bernstein03} on 27-29 July 2021 with the 0\farcs5 slit and 2x1 binning, providing resolution $R \approx 50,000/40,000$ on the blue/red arm of MIKE, respectively that span 3300-5000 \AA\ and 4900-10,000 \AA. These stars were selected based on their brightness and observability from Magellan from those satisfying Eqns. \ref{eqn:kraken} and \ref{eqn:gse}. The data were reduced with CarPy \citep{Kelson03}.
The spectra were analyzed using SMHR\footnote{\url{https://github.com/andycasey/smhr}, first described in \citealt{Casey14}}, which provides an interface to doppler correct, normalize and stitch orders, fit equivalent widths, interpolate \citet{Castelli04} stellar atmospheres, and run MOOG including scattering opacity \citep{Sneden73,Sobeck11} and \citet{Barklem00} damping\footnote{\url{https://github.com/alexji/moog17scat}} to determine abundances from equivalent widths and spectrum synthesis (see \citealt{Ji20b} for a detailed description).

The analyzed lines were selected primarily from \citet{Jonsson2017a}, \citet{Lomaeva2019}, and \citet{Forsberg2019}, who selected good unblended lines for metal-rich red giant stars in the disk and bulge.
We supplemented these lists with lines selected by \citet{Roederer2018}, both to measure more elements and to replace lines that were undetected in any of our stars.
For Fe I, Fe II, O I, Mg I, Ca I, and Ti I, we adopted the atomic data from \citet{Jonsson2017a} when available to ensure our stellar parameters and $\alpha$-abundances are on the same scale as their results.
The atomic data for other species (Na I, Al I, Si I, K I, Sc II, V I, Mn I, Ni I, Zn I, Y II, Ba II, La II, Ce II, Eu II) were taken from \code{linemake} \citep{Placco2021}, which keeps up-to-date libraries of experimentally measured oscillator strengths.

Stellar parameters were determined by balancing excitation potential vs. Fe I abundance for \Teff, ionization balance for Fe I and II for \logg, line strength vs. Fe I abundance for $\nu_t$, and setting the model metallicity to [Fe I/H]. We adopted [$\alpha$/Fe]$=+0.4$ atmospheres for the analysis, but we verified that the results and conclusions are unchanged for [$\alpha$/Fe]$=0.0$.
Since we used the same Fe lines as \citet{Jonsson2017a}, we adopted their systematic uncertainties of 50 K, 0.15 dex, 0.10 \kms, and 0.05 dex for \Teff, \logg, $\nu_t$, and [M/H] respectively.
Statistical uncertainties were determined using the standard error on the respective slopes or abundance differences.
Stellar parameter uncertainties were propagated to abundance uncertainties following \citet{Ji20b}.

\begin{figure*}
\centering
\includegraphics[width=\linewidth]{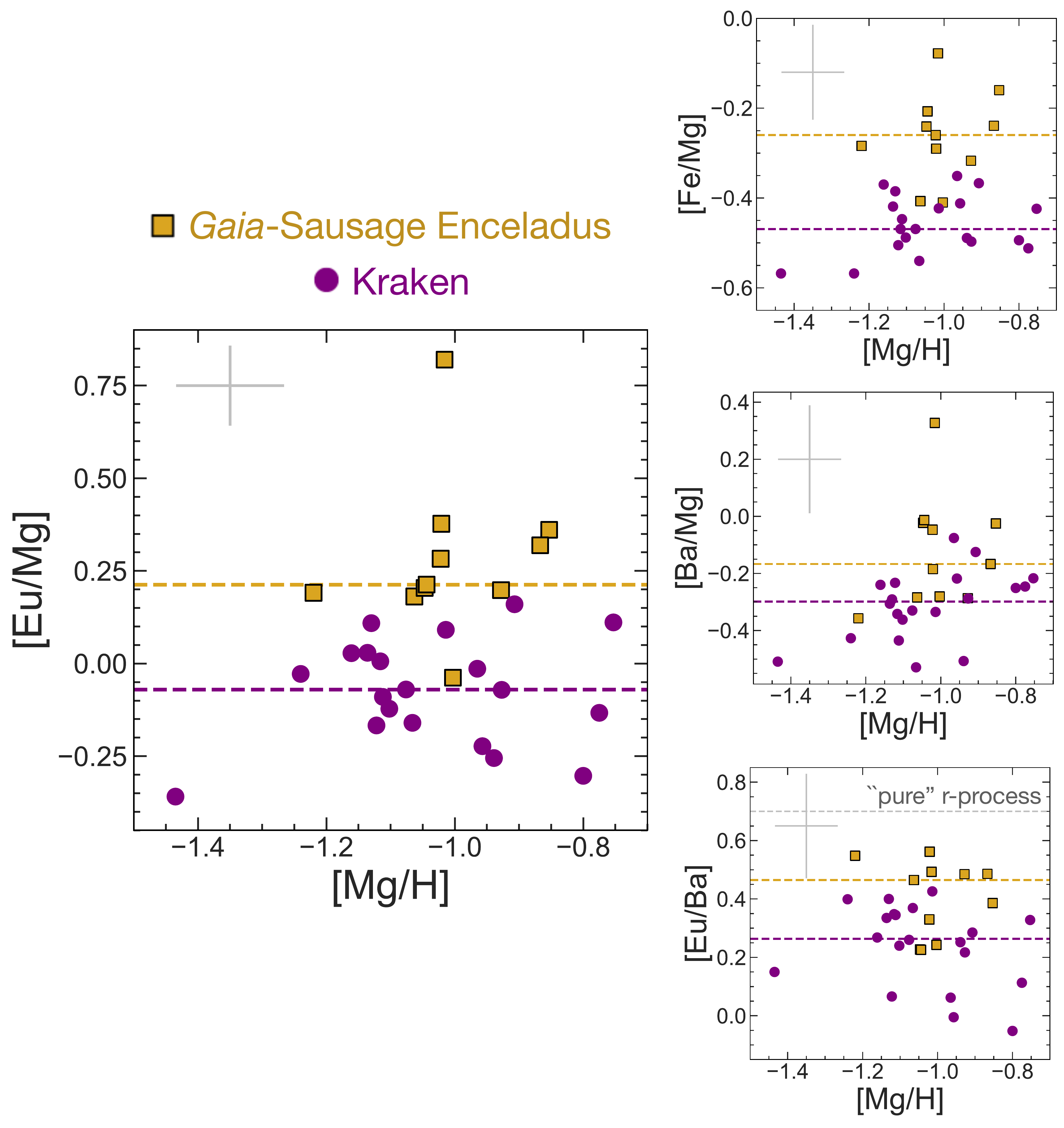}
\caption{Magellan/MIKE abundances for GSE (golden squares) and Kraken (purple circles). Each panel shows the median abundances as dashed lines. Typical errors are indicated in the top-left corner. Notice that Kraken has a relative dearth of elements that are associated with ``delayed" sources (e.g., Eu: NSMs, Fe: Type Ia SNe, Ba: NSMs+AGBs) across all panels, supporting the case for a more efficient, shorter star-formation history compared to GSE.}
\label{fig:MIKE}
\end{figure*}

\section{Mass \& star-formation duration}
\label{sec:massandsfh}

Here we discuss two fundamental parameters of Kraken and GSE -- mass and star-formation duration -- that are crucial to interpreting their chemical evolution.

Both the stellar mass and halo mass of GSE and Kraken have been found to be similar via a variety of methods. For instance, comparing the chemodynamical properties of the accompanying GC systems against a suite of hydrodynamical MW simulations, \citet[][]{Kruijssen20} inferred $M_{\star}=2.7^{+1.1}_{-0.8}\times 10^{8} M_{\rm{\odot}}$, $M_{\rm{halo}}= 9.6^{+1.6}_{-1.7} \times 10^{10} M_{\rm{\odot}}$ for GSE, and $M_{\star}=1.9^{+1.0}_{-0.6}\times 10^{8} M_{\rm{\odot}}$, $M_{\rm{halo}}= 8.3^{+2.2}_{-1.7} \times 10^{10} M_{\rm{\odot}}$ for Kraken. Other methods yield consistent results and include GC to halo mass relations (GSE \& Kraken: \citealt[][]{Forbes20}), chemical evolution models (GSE: \citealt[][]{Fernandez-Alvar18}, Kraken: \citealt[][]{Horta21}), tailored N-body simulations (GSE: \citealt[][]{Naidu21}), halo star counts (GSE: \citealt[][]{Mackereth20}), and the mass metallicity relation (assuming high-$z$ evolution of the relation from \citealt[][]{Ma16MZR}, and accretion redshifts from \citealt[][]{Bonaca20} and \citealt[][]{Kruijssen20}).

Since GSE comprises the bulk of halo stars in the Solar neighborhood, ages for hundreds of GSE main-sequence turnoff stars have been measured via high-resolution spectroscopy combined with optical through IR photometry \citep[][]{Bonaca20}. The age distributions show virtually all GSE stars to be $>10.2$ Gyrs old, with an SF duration of $3.6^{+0.1}_{-0.2}$ Gyrs.

For Kraken, which is buried in the dusty Galactic center and has therefore been studied more sparsely, we rely on less direct tracers. A lower-limit comes from the age-spread of the GCs associated with Kraken \citep[][]{Massari19,Kruijssen20} -- the GC age-spread is 1 Gyr as per \citet{Dotter10,Dotter11} and 1.5 Gyr as per \citet{VandenBerg13}. An upper limit comes from observing that Kraken must have been disrupted before GSE owing to its depth in the potential and truncated chemical sequence with no ``knee" in the [Fe/H] vs. [Mg/Fe] plane (Figure \ref{fig:krakensel}), i.e., Kraken's SF duration must be $<3.6$ Gyr. As our fiducial value we adopt 2 Gyr based on \citet{Kruijssen20} who used all available information on the Kraken GCs (age, metallicity, dynamics) to infer $\approx1.7$ Gyrs as the difference between GSE and Kraken's accretion epochs.

The key takeaway from this section is that the similar stellar and halo masses of Kraken and GSE make for a controlled experiment, wherein observed differences in chemistry must arise mainly due to the differing star-formation durations. Almost all physical processes important to chemical evolution are to first order a function of halo mass and the depth of the potential, and must thus be similar in both galaxies (e.g., the fraction of enriched gas lost to outflows, neutron stars that become unbound due to natal kicks).
This contrasts with previous studies of $r$-process enrichment in intact dwarf galaxies \citep[e.g.,][]{Duggan2018,Skuladottir20}, which by necessity compare galaxies of very different stellar and halo masses and are more susceptible to galaxy formation uncertainties.

\section{Abundance Results}
\label{sec:results}

In Figure \ref{fig:MIKE} we contrast abundances of Eu, Mg, Fe, and Ba measured for Kraken and GSE. The Kraken sample spans [Fe/H] of -2.0 to -1.2 (median: $-1.52^{+0.08}_{-0.03}$) whereas the GSE sample spans [Fe/H] of -1.5 to -1.0 (median: $-1.28^{+0.03}_{-0.03}$). Median abundances are indicated with a dashed line in all panels. We plot and analyze abundances with respect to Mg instead of Fe since Mg has a single production channel (CCSNe) that is tightly coupled to star formation, and thus greatly simplifies our interpretation. We stress again that GSE and Kraken have similar stellar and halo masses, so they make for a controlled setting to reveal the effects of differing star-formation durations ($\approx2$ Gyr vs. 3.6 Gyr) on chemical evolution.

The key empirical result of this paper is highlighted in the left panel of Figure \ref{fig:MIKE} -- at similar [Mg/H], GSE has an elevated $\langle$[Eu/Mg]$\rangle$ compared to Kraken by $\approx0.3$ dex (median [Eu/Mg] of $0.22^{+0.10}_{-0.02}$ and $-0.07^{+0.06}_{-0.04}$ respectively). Mg is exclusively produced by CCSNe and thus closely tracks the star-formation history (SFH), whereas Eu is produced in the $r$-process. 
Since GSE and Kraken had similar stellar masses, at fixed [Mg/H] they were enriched by a similar number of CCSNe. Thus,
the only reason for higher [Eu/Mg] in GSE is its $\approx2\times$ longer star-formation duration, which is strong evidence for a delayed channel of Eu production (e.g., NSMs). We note that the [Eu/Mg] distribution of GSE we measure is consistent with studies based on the GALAH Survey \citep[][]{Matsuno21,Buder21} and a study of [Fe/H]$<-1.5$ GSE stars \citep[][]{Aguado20}.

Eu is almost exclusively produced via the $r$-process, which may have two channels -- an instant channel that tracks the star-formation linked to rCCSNe and a delayed channel due to NSMs. At fixed [Mg/H], the number of CCSNe (and rCCSNe) is controlled for, and so a significantly higher [Eu/Mg] in GSE implies it has been enriched to a far greater extent by NSMs. This striking difference due to NSMs across $\approx2$ Gyr (Kraken) and $3.6$ Gyr (GSE) star-formation durations is strong evidence that enrichment from rCCSNe+NSMs is still evolving $>2$ Gyrs after the onset of star-formation in these systems. The magnitude and speed of this evolution provide strong constraints on proposed channels of the $r$-process that we explore in \S\ref{sec:models}.

The panels in Figure \ref{fig:MIKE} depicting a higher [Fe/Mg] and [Ba/Mg] at fixed [Mg/H] further support the overall picture that GSE had a more extended, less efficient star-formation history compared to Kraken. GSE is more enriched in elements that are expected to arise from delayed channels (e.g., Ba from asymptotic giant branch stars (AGBs) \& NSMs, Fe from Type Ia SNe) for a similar number of CCSNe as Kraken.

In the final panel of Figure \ref{fig:MIKE} we show [Eu/Ba], which is used as a measure of the relative contributions from the $r$-process and s-process, the dominant production pathways for Eu and Ba respectively. For pure $r$-process enrichment, the expected [Eu/Ba] is $\approx0.7$ \citep{Sneden2008}. More than half the GSE sample lies close to this limit, whereas Kraken stars have lower [Eu/Ba] and span a wider range ($\approx-0.1$ to $\approx0.4$). The $r$-process in GSE is prolific enough to propel stars closer to the ``pure" $r$-process limit. In what follows we focus on [Eu/Mg], and defer the analysis of the remaining abundances (e.g., Ba) that are of great interest for a variety of issues pertaining to chemical evolution to future work.

\section{Interpreting [E\lowercase{u}/M\lowercase{g}] with Simple Chemical Evolution Models}
\label{sec:models}

Because we have good estimates for the star-formation durations of GSE and Kraken (\S\ref{sec:massandsfh}), it should be possible to infer a typical timescale for the delayed $r$-process production.
To illustrate the power of galaxies with different formation timescales but same final mass, here we produce simple models to translate the difference in [Eu/Mg] to constraints on the $r$-process production channels. We emphasize that these models are meant to give an illustrative understanding of the key factors at play, but more complex models will be needed for quantitative constraints using these abundances \citep[e.g.,][]{Molero21}.

\begin{figure*}[t]
\centering
\includegraphics[width=\linewidth]{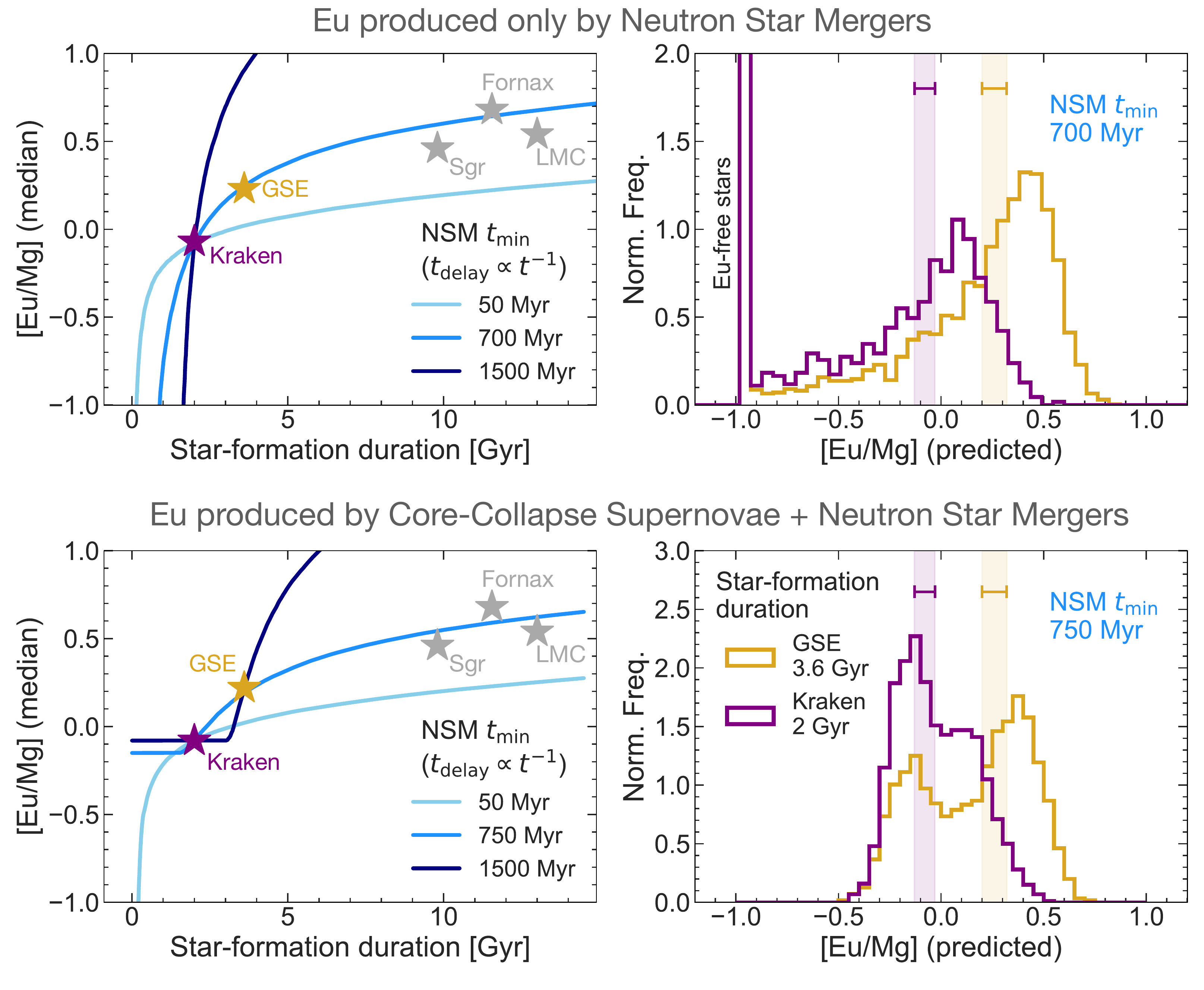}
\caption{Simple models for the evolution of [Eu/Mg]. Each track shown in blue is constrained to pass through the Kraken point (purple star) by adjusting $R_{\rm{CCSNe}}/R_{\rm{NSMs}}$ and the [Eu/Mg]$_{\rm{CCSNe}}$. On these tracks every point corresponds to an $M_{\rm{\star}} = 5\times10^{8} M_{\rm{\odot}}$ galaxy. The predicted [Eu/Mg] distributions shown on the right include a 0.1 dex uncertainty for comparison with observations. The median [Eu/Mg] we measure is shown as shaded regions, and all Eu-free stars are set to [Eu/Mg]$=-1$. Statistical errors on the median [Eu/Mg] for all galaxies shown in the left panels are smaller than the star-shaped markers used to represent them. \textbf{Top:} Models where Eu arises purely from NSMs require minimum delay times $\approx500-1000$ Myrs to match the rapid evolution across Kraken and GSE (top-left). Such delay times imply the existence of substantial fractions of Eu-free stars inconsistent with observations (top-right). \textbf{Bottom:} When contributions from both CCSNe and NSMs are allowed, there is a ``floor" for [Eu/Mg] at early times due to CCSNe followed by a rapid rise from delayed NSMs. Models with $t_{\rm{min}}\approx500-1000$ Myrs reproduce the abundances in Kraken and GSE as well as in dwarfs with extended star-formation histories (Sgr, LMC, Fornax). The predicted [Eu/Mg] distribution (right panel) is bimodal, with each mode corresponding to the CCSNe phase and NSM plateau phase of the tracks.}
\label{fig:models}
\end{figure*}

\subsection{Setup}
We model NSMs that result from a star-formation episode with a delay time distribution (DTD, $t_{\rm{delay}}\propto t^{-1}$) -- no NSMs occur before the mininum delay time ($t_{\rm{min}}$). This choice is motivated by theoretical population synthesis results \citep[e.g.,][]{Neijssel19} and observations that suggest short gamma ray bursts (likely tracing NSMs) follow a DTD of the same shape as Type Ia supernovae \citep[e.g.,][]{Paterson20}. The Eu yield of NSMs is fixed to $M_{\rm{Eu}} = 10^{-4.5} M_{\rm{\odot}}$ per NSM, inferred from the Eu-enhanced ultra-faint dwarfs \citep{Ji16} and consistent with GW170817 \citep{Cote18}.

For CCSNe and rCCSNe we assume Eu and Mg production occur simultaneously, immediately after a starburst. We adopt a \citet[][]{Kroupa01} initial mass function with a cutoff of 300 $M_{\rm{\odot}}$ and assume every star $>10 M_{\rm{\odot}}$ ends its life as a CCSNe. The effective Eu yield averaged over all CCSNe ($\rm{[Eu/Mg]}_{\rm{CCSNe}}$) is left as a free parameter and assumed to be unchanging with time (i.e., the fraction and yield of rCCSNe relative to CCSNe is fixed). The relative rate, $\log(R_{\rm{CCSNe}}/R_{\rm{NSMs}})$, is allowed to vary from 2-4 based on estimates for NSMs from gravitational wave observations ($286^{+510}_{-237}$ Gpc$^{-3}$ yr$^{-1}$, \citealt[][]{Abbott21NSMrate}) and for CCSNe from transient surveys ($1.01^{+0.50}_{-0.35}\times10^{5}$ Gpc$^{-3}$ yr$^{-1}$, \citealt[][]{Perley20}). Note that these are \textit{local} estimates, whereas for Kraken and GSE the $z>2$ rates are of interest -- however, our wide adopted range on the rates likely encompasses the mild evolution expected with redshift \citep[e.g.,][]{Neijssel19,Santoliquido21}.

We set the GSE and Kraken star-formation histories to be constant in time (``top-hat") motivated by the observed GSE age distribution \citep[][]{Bonaca20}. We assume instant mixing such that enriched gas is converted to stars with no delay -- in practice we expect inhomogenous mixing to produce scatter in abundances around the mean trends we predict. With these ingredients we are able to produce tracks of [Eu/Mg] as a function of star-formation duration. Our simple models are able to perfectly reproduce the more sophisticated [Eu/Mg] chemical evolution model for GSE by \citet{Matsuno21}.

In addition to the [Eu/Mg] data we have measured for Kraken and GSE, we also compare our models qualitatively against [Eu/Mg] measured for dwarf galaxies of comparable mass (LMC, Fornax, and Sgr). For the LMC we draw on \citet[][]{vanderswaelmen13} who measured [Eu/Mg] for 94 stars spanning [Fe/H] of -0.9 to -0.4 (10$^{\rm{th}}$ and 90$^{\rm{th}}$ percentiles, median [Fe/H] of $-0.66^{+0.02}_{-0.02}$). The [Eu/Mg] for Fornax is from \citet[][]{Letarte18}, who present [Eu/Mg] for 70 stars with [Fe/H] spanning -1.5 to -0.6 (median [Fe/H] of $-0.86^{+0.02}_{-0.01}$). For Sgr, we rely on \citet[][]{Bonifacio00,McWilliam13} who measured these abundances for five stars spanning [Fe/H] of -0.1 and -0.5. Star-formation durations for these galaxies (all $\gtrsim10$ Gyrs) are sourced from \citet[][]{Weisz14sfh} -- in particular, we adopt their $t_{\rm{90}}$ which is the time taken to form 90$\%$ of the stellar mass. These galaxies do not have simple top-hat star-formation histories, their abundances are measured on different scales, and [Eu/Mg] is measured mostly at the metal-rich end of these galaxies, i.e., they are probing the [Eu/Mg] at the very end of the SFH and not the median [Eu/Mg] across the entire SFH. For these reasons we do not make detailed quantitative comparisons, but nonetheless plot these data as indicative of the long-run ``plateau" [Eu/Mg] in galaxies with extended star-formation histories.

\subsection{Model I: only rCCSNe produce Eu}

The first model we briefly consider assumes Eu is produced solely by rCCSNe. Such a model is already disfavored based on Figure \ref{fig:MIKE}, since GSE shows higher [Eu/Mg] than Kraken at similar [Mg/H] (i.e., similar number of CCSNe and rCCSNe). As per our setup, rCCSNe produce flat tracks in [Eu/Mg] as a function of star-formation duration since both Eu and Mg are produced in similar proportions over time. For a significant difference in [Eu/Mg] to emerge across Kraken and GSE time-dependent/delayed sources that increase Eu production with time are necessary.

\subsection{Model II: only NSMs produce Eu}

In this model we assume NSMs are solely responsible for all Eu production. When NSMs are the only $r$-process channel, $t_{\rm{min}}$ (the minimum delay time) is the key parameter that sets the relative [Eu/Mg]. We show tracks spanning a range of $t_{\rm{min}}$ in the top-left panel of Figure \ref{fig:models} constrained to match the Kraken value. Short $t_{\rm{min}}$ (e.g., 10-50 Myr) result in rapidly plateauing [Eu/Mg] at odds with the data. We find $t_{\rm{min}}=611^{+147}_{-151}$ Myrs by adopting a uniform prior between 0-3.6 Gyr and maximizing the likelihood such that there is a difference in median [Eu/Mg] of $0.30\pm0.05$ dex across the Kraken and GSE SF durations.

However, a $>500$ Myr delay time implies the existence of substantial fractions of Eu-free stars that formed prior to the onset of NSMs (top-right, Figure \ref{fig:models}). At least $\approx25\%$ of the stars in Kraken and $\approx15\%$ of the stars in GSE would be Eu-free. We do not observe any Eu-free stars. For GSE this might be because our sample does not extend below [Fe/H]$<-1.5$ to probe ages $\lesssim500$ Myr ($\approx15\%$ of the SFH). Though \citet[][]{Aguado20} find Eu detections ([Eu/Mg]$\approx0$) in all four GSE stars they studied with [Fe/H] of $-1.4$ to $-1.8$. More strikingly, in Kraken, for which half our sample is at [Fe/H]$<-1.5$, we detect Eu in every star, including in stars as metal-poor as [Fe/H]$=-2.0$. Note that assuming a steeper delay time distribution ($t_{\rm{delay}}\propto t^{-1.5}$, e.g., \citealt[][]{Cote19}) only makes matters worse by favoring longer $t_{\rm{min}}$ as the [Eu/Mg] plateau occurs even more rapidly. On the other hand a much shallower $t_{\rm{delay}}\propto t^{+0.5}$ \citep[][]{Tsujimoto21} produces a gently rising [Eu/Mg], with no Eu-free stars, but is disfavored by NSM models\footnote{In detail, \citet[][]{Tsujimoto21} are effectively fitting the \textit{enrichment} DTD as $t_{\rm{delay}}\propto t^{+0.5}$ and not the \textit{merger} DTD. See \S\ref{sec:kicks} for discussion of the distinction. Accounting for this, our findings are in excellent qualitative agreement (i.e., NSMs+rCCSNe are required, NSMs must produce delayed enrichment).} \citep[e.g.,][]{Chruslinska18,Belczynski18,Neijssel19}.

\subsection{Model III: rCCSNe+NSMs produce Eu}

In this model we assume Eu arises from rCCSNe as well as NSMs. The dearth of Eu-free stars that emerges from Model II can be explained by contributions from early rCCSNe. rCCSNe produce Eu promptly after star-formation, and therefore ensure a floor for [Eu/Mg] at early times. Once the NSMs begin contributing after the minimum delay time, $t_{\rm{min}}$, the [Eu/Mg] begins rising.

For a floor of [Eu/Mg]$\gtrsim -0.25$ suggested by the Kraken distribution in Figure \ref{fig:MIKE}, we find only models with $t_{\rm{delay}}\gtrsim500$ Myr are able to match the evolution across Kraken and GSE (bottom row, Figure \ref{fig:models}). Requiring a plateau value of [Eu/Mg]$\approx0.5$ for $\approx10$ Gyr star-formation durations observed in galaxies of comparable mass -- Sagittarius, Fornax, and the Large Magellanic Cloud -- provides an upper bound of $t_{\rm{delay}}\lesssim1000$ Myr (see navy blue curve in bottom-left panel of Figure \ref{fig:models}). The predicted [Eu/Mg] distributions for GSE and Kraken are consistent with the data at hand -- the distributions are bimodal, with one mode corresponding to the rCCSNe-only phase and the other close to the peak [Eu/Mg] reached during the rCCSNe+NSM phase.

\begin{figure*}[t]
\centering
\includegraphics[width=\linewidth]{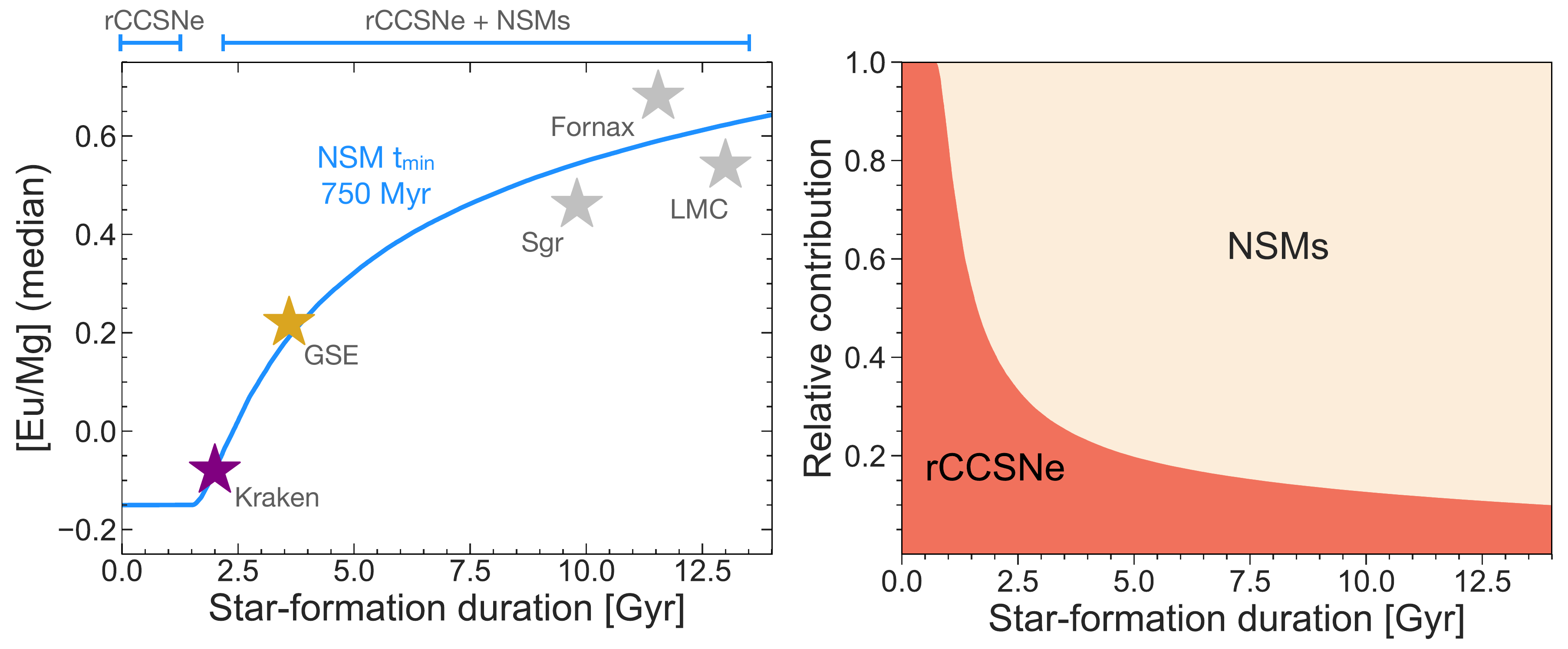}
\caption{\textbf{Left:} Schematic of our preferred scenario featuring both rare CCSNe and NSMs with a $t_{\rm{min}}=750$ Myr. Only systems with stellar masses comparable to Kraken and GSE are shown for fair comparison. \textbf{Right:} The relative contribution of rCCSNE and NSMs to the total Eu produced as a function of star-formation duration. rCCSNe contribute significantly at early times in galaxies like Kraken ($\approx50\%$ of Eu), whereas in galaxies like Sgr and Fornax $\approx85\%$ of Eu arises from NSMs.}
\label{fig:schematic}
\end{figure*}

\newpage
\section{Discussion}
\label{sec:discussion}

\subsection{A role for both rCCSNe and NSMs}
\label{sec:bothCCSNeNSM}
Our models show both rCCSNe and NSMs have a role to play in $r$-process enrichment -- rCCSNe dominate at early times ($\lesssim2$ Gyr), while NSMs take over at later epochs (Figure \ref{fig:schematic}). Without delayed enrichment from NSMs, the sharp rise in [Eu/Mg] with star-formation duration is not possible, and without rCCSNe a large population of Eu-free stars would be produced at early times. For our fiducial model shown in the left-panel of Figure \ref{fig:schematic}, $\approx50\%$ of all Eu produced in Kraken arises from rCCSNe. However, for longer star-formation durations (e.g., in Sgr), NSMs dominate and account for $\approx85\%$ of the Eu. Note that the yield of individual rCCSNe (e.g., collapsars) producing Eu could be much higher than NSMs \citep[e.g.,][]{Siegel19} but the effective yield averaged over all CCSNe (i.e., accounting for the rarity of rCCSNe) is lower than NSMs (see right panel of Figure \ref{fig:schematic}).

The idea that both rCCSNe and NSMs are required to explain $r$-process in galaxies ranging from the MW to dwarf galaxies is not new.
Several recent models have used both sites to successfully reproduce $r$-process chemical evolution in the MW, especially if CCSNe stop producing $r$-process at higher metallicities \citep[e.g.,][]{Ting12,Matteucci14, Cescutti15, Wehmeyer15, Haynes19, Cote19, Siegel19}.
The latter is expected in currently viable $r$-process mechanisms for rCCSNe, because high angular momentum is needed to create jets or accretion disks \citep[e.g.,][]{Mosta18, Siegel19} and metal-rich stars lose a larger fraction of their angular momentum to winds. It also appears likely that both rCCSNe and NSMs are needed to explain Eu evolution of intact dwarf galaxies with $M_\star$ ranging from $10^{4-8} M_\odot$ \citep[e.g.,][]{Ji16c,Skuladottir19,Molero21,delosReyes21}. However, we note that all chemical evolution models, including our simple models here, can only show that the combination of rCCSNe and NSMs is sufficient, but not necessary, to explain the data. The crucial uncertainty is whether very prompt NSMs (${\lesssim}10$ Myrs) that are able to enrich star-forming regions are sufficiently common because these NSMs would effectively mimic rCCSNe from a chemical evolution standpoint \citep[e.g.,][]{Beniamini19,Safarzadeh19,Simonetti19,Andrews20,Romero-Shaw20,Kirby20}. However, there is some evidence against fast mergers being a dominant population from short gamma ray burst redshift distributions and host galaxy populations \citep[e.g.,][]{Cote19,Simonetti19}, so the most likely scenario appears to be that both rCCSNe and NSMs are required.

\subsection{Natal kicks and cooling times may explain the $\approx500-1000$ Myr delay time for enrichment from NSMs}
\label{sec:kicks}

At face value, our models suggest a minimum delay time, $t_{\rm{min}}\approx500-1000$ Myr for NSMs, seemingly at odds with the vast majority of theoretical binary evolution models that predict $t_{\rm{min}}$ on the order of $\approx10-30$ Myrs \citep[e.g.,][]{Chruslinska18,Belczynski18,Neijssel19}. However, our inferred $t_{\rm{min}}$ not only includes the time it takes for neutron stars to merge, but also the time required for the produced elements to find their way into subsequent generations of stars. That is, the delay time for mergers, and the delay time for \textit{enrichment} from mergers could be substantially different.

This substantial difference -- a $\approx10-30$ Myr delay time for mergers, but a $500-1000$ Myr delay for $r$-process enrichment due to NSMs -- may be explained by considering natal kicks (velocity impulses from supernova explosions) from NSMs. Due to both these mechanisms, NSMs potentially enrich gas far from active star-forming regions of their hosts, thus altering patterns of $r$-process enrichment \citep[][]{Bramante2016,Macias19,Banerjee20,vandevoort21}. Natal kicks propel neutron stars off the disk -- short GRBs (likely tracing NSMs) are observed at typical distances of $\approx1.5\times$ half-light radii, and often off the plane, in $10^{8.5}-10^{11.5} M_{\rm{\odot}}$ galaxies \citep[e.g.,][]{Fong13}. This effect is likely more pronounced in $z\gtrsim2$ systems like GSE and Kraken, whose disks are thick, turbulent, and have relatively lower binding energies versus the more settled disks seen in local, $z\approx0$ systems \citep[e.g.,][]{Bird13,Ma17b,Park20}.

Due to both these effects, the NSMs in GSE and Kraken may have merged with $t_{\rm{min}}\approx10-30$ Myrs, but may have ended up enriching gas outside the interstellar medium (ISM) that was yet to cool into a star-forming state\footnote{\citet{Schonrich19b} considered a multi-phase ISM for the MW and found the opposite conclusion, that NSMs would have to rapidly enrich cold ISM instead of hot ISM; however, they only considered $r$-process production in NSMs without CCSNe.}. The expected cooling/condensation time for extra-planar gas at $\approx1-2 r_{\rm{e}}$ for Kraken and GSE-like galaxies at $z\approx2-3$ is expected to be on the order of few hundred Myrs to a Gyr (Eqn. 12 of \citealt[][]{Fraternali17}, Eqns. 8.94-8.95 of \citealt[][]{MvdBW}; GSE physical parameters from \citealt[][]{Naidu21}), comparable to the dynamical time of the halos. Thus, the cooling time may entirely account for the high $t_{\rm{min}}$ we infer. 

The implication is that metal poor stars that probe the first few hundred Myrs of star-formation in a variety of systems, and that are widely analyzed to understand the $r$-process may be exclusively sampling yields from rCCSNe since the $r$-process enriched gas is yet to rain down and form stars. This also explains why galaxies like Sculptor, Sagittarius, and Fornax have relatively flat [Eu/Mg]$\approx-0.1$ to 0.0 sequences as a function of [Fe/H] and age \citep[][]{Skuladottir20}, during the first few Gyrs of their star-formation history when rCCSNE ostensibly account for their $r$-enrichment (right panel of Figure \ref{fig:schematic}).

\subsection{Caveats \& Outlook}
\label{sec:caveats}

A persisting puzzle entirely independent of the results presented here is why the MW  has a gently evolving [Eu/Mg]$\approx0.0-0.1$ for stars that formed $0-10$ Gyrs ago, and a virtually flat track in [Eu/Mg] vs. [Fe/H] with almost no stars reaching [Eu/Mg]$\approx0.5$ \citep[][]{Skuladottir20}. Galaxies like the LMC, Fornax, and Sgr have much higher [Eu/Mg] at fixed metallicity compared to the MW. Our simple rCCSNe+NSM model (Figure \ref{fig:schematic}) explains the [Eu/Mg] evolution in these galaxies, but not in the MW. Sophisticated models that account for the MW's rich merger history, complex SFH, differential mixing, and inside-out growth are required to understand the Galactic [Eu/Mg] evolution.

We emphasize once again that our simple analytical models are meant to give a broad, qualitative sense for the situation. There are additional complexities that are well-motivated, but poorly empirically constrained, that we do not explore here -- e.g., metallicity-dependent $r$-process yields, deviations from simple $t^{-1}$ DTDs, departures from the instant enrichment assumption.

We adopted a star-formation duration of 2 Gyrs for Kraken based on \citet[][]{Kruijssen20}, but the conservative range for this quantity is $\approx1.5-3.6$ Gyrs (\S\ref{sec:massandsfh}). Age distributions from Kraken's MSTO stars and self-consistent age determinations for the entire sample of Kraken GCs (only half have published ages) will tighten this range, and also test the validity of our top-hat star-formation history assumption. For now, we note that $>2$ Gyr ($<2$ Gyr) star-formation durations produce longer (shorter) $t_{\rm{min}}$ for enrichment from NSMs that are still $\approx10-100\times$ larger than the $t_{\rm{min}}\approx10-30$ Myr expected from theory. For instance, a $>1.5$ Gyr star-formation duration for Kraken results in a $>300$ Myr delay for Models II and III.

Future observations can further test our proposed picture by analyzing additional disrupted dwarfs. Of particular interest are systems like I'itoi \citep[][]{Naidu20} and Thamnos \citep[][]{Koppelman19} which may have star-formation durations even shorter than Kraken and may thus directly constrain the pure Eu yield of rCCSNe. For Kraken and GSE themselves, abundances for larger samples spanning the entirety of their star-formation histories will enable detailed comparisons against the predicted bimodal [Eu/Mg] distributions from our preferred model. Confirming the existence and location of these modes will yield rich insights.

The convenient, star-by-star access to ``high-$z$" galaxies afforded by the disrupted dwarfs within a few kpc from the Sun is set to transform our understanding of early Universe chemistry as more of these systems are unearthed and characterized. This work provides a glimpse of the unique constraints that might be possible with dozens of such systems in the imminent future.

\facilities{Magellan (MIKE), \textit{Gaia}}

\software{
    \package{IPython} \citep{ipython},
    \package{matplotlib} \citep{matplotlib},
    \package{numpy} \citep{numpy},
    \package{scipy} \citep{scipy},
    \package{jupyter} \citep{jupyter},
    \package{gala} \citep{gala1, gala2},
    \package{Astropy}
    \citep{astropy1, astropy2},
    \package{smhr} \citep{Casey14},
    \package{CarPy} \citep{Kelson03},
    \package{MOOG} \citep{Sneden73,Sobeck11}
    }
    
\acknowledgments{We thank ace Magellan observer Yuri Beletsky for collecting these data for us amidst the vicissitudes of a pandemic. We are grateful to the CfA and U. Chicago TACs for their continued support of this long-term project. We thank Tadafumi Matsuno and Yutaka Hirai for sharing their GSE chemical evolution tracks from \citet[][]{Matsuno21} that gave us great confidence in our elementary models. We thank Anna-Christina Eilers for sharing an updated version of the \citet[][]{Hogg19,Eilers19} catalog with us. This project was inspired by a renegade meeting on the sidelines of EAS 2021 whose ringleaders were Ana Bonaca, Chervin Laporte, and Diederik Kruijssen. We acknowledge illuminating conversations with Aaron Dotter and Seth Gossage on the Kraken globular clusters, the kindness of Holger Baumgardt in helping compile their CMDs, and the generosity of Christian Johnson for discussing the many mysteries of NGC 6273. We had the fortune of discussing NSM kicks with Freeke van de Voort. RPN thanks Michelle Peters for her infinite kindness on post-observing days. 

RPN acknowledges an Ashford Fellowship granted by Harvard University. CC acknowledges funding from the Packard foundation. YST acknowledges financial support from the Australian Research Council through DECRA Fellowship DE220101520.

This work has made use of data from the European Space Agency (ESA) mission
{\it Gaia} (\url{https://www.cosmos.esa.int/gaia}), processed by the {\it Gaia}
Data Processing and Analysis Consortium (DPAC,
\url{https://www.cosmos.esa.int/web/gaia/dpac/consortium}) \citep{dr2ack1,GaiaEDR3}. Funding for the DPAC
has been provided by national institutions, in particular the institutions
participating in the {\it Gaia} Multilateral Agreement. 

Funding for the Sloan Digital Sky Survey IV has been provided by the 
Alfred P. Sloan Foundation, the U.S. 
Department of Energy Office of 
Science, and the Participating 
Institutions. SDSS-IV acknowledges support and 
resources from the Center for High 
Performance Computing  at the 
University of Utah. The SDSS 
website is www.sdss.org. SDSS-IV is managed by the Astrophysical Research Consortium 
for the Participating Institutions 
of the SDSS Collaboration including 
the Brazilian Participation Group, 
the Carnegie Institution for Science, 
Carnegie Mellon University, Center for 
Astrophysics | Harvard \& 
Smithsonian, the Chilean Participation 
Group, the French Participation Group, 
Instituto de Astrof\'isica de 
Canarias, The Johns Hopkins 
University, Kavli Institute for the 
Physics and Mathematics of the 
Universe (IPMU) / University of 
Tokyo, the Korean Participation Group, 
Lawrence Berkeley National Laboratory, 
Leibniz Institut f\"ur Astrophysik 
Potsdam (AIP),  Max-Planck-Institut 
f\"ur Astronomie (MPIA Heidelberg), 
Max-Planck-Institut f\"ur 
Astrophysik (MPA Garching), 
Max-Planck-Institut f\"ur 
Extraterrestrische Physik (MPE), 
National Astronomical Observatories of 
China, New Mexico State University, 
New York University, University of 
Notre Dame, Observat\'ario 
Nacional / MCTI, The Ohio State 
University, Pennsylvania State 
University, Shanghai 
Astronomical Observatory, United 
Kingdom Participation Group, 
Universidad Nacional Aut\'onoma 
de M\'exico, University of Arizona, 
University of Colorado Boulder, 
University of Oxford, University of 
Portsmouth, University of Utah, 
University of Virginia, University 
of Washington, University of 
Wisconsin, Vanderbilt University, 
and Yale University.}

\bibliography{MasterBiblio}
\bibliographystyle{apj}

\end{CJK*}
\end{document}